\documentclass[conference]{IEEEtran}
\pdfoutput=1

\usepackage[colorinlistoftodos,prependcaption,textsize=tiny]{todonotes}

\usepackage{url}

\usepackage{amsmath,amssymb,amsfonts}
\usepackage{lmodern}
\usepackage{float}
\usepackage{enumitem}
\usepackage[utf8]{inputenc}
\DeclareUnicodeCharacter{2010}{-}
\setlist[itemize]{leftmargin=*}
\usepackage{stfloats}

\usepackage{xcolor}
\usepackage{listings}
\definecolor{mGreen}{rgb}{0,0.6,0}
\definecolor{mGray}{rgb}{0.5,0.5,0.5}
\definecolor{mPurple}{rgb}{0.58,0,0.82}
\definecolor{backgroundColour}{rgb}{0.95,0.95,0.92}
\lstdefinestyle{CStyle}{
    backgroundcolor=\color{backgroundColour},   
    commentstyle=\color{mGreen},
    keywordstyle=\color{magenta},
    numberstyle=\tiny\color{mGray},
    stringstyle=\color{mPurple},
    basicstyle=\footnotesize,
    breakatwhitespace=false,         
    breaklines=true,                 
    captionpos=b,                    
    keepspaces=true,                 
    numbers=left,                    
    numbersep=5pt,                  
    showspaces=false,                
    showstringspaces=false,
    showtabs=false,                  
    tabsize=2,
    language=C
}

\usepackage{scalerel}
\usepackage{tikz}
\usetikzlibrary{svg.path}
\definecolor{orcidlogocol}{HTML}{A6CE39}
\tikzset{
  orcidlogo/.pic={
    \fill[orcidlogocol] svg{M256,128c0,70.7-57.3,128-128,128C57.3,256,0,198.7,0,128C0,57.3,57.3,0,128,0C198.7,0,256,57.3,256,128z};
    \fill[white] svg{M86.3,186.2H70.9V79.1h15.4v48.4V186.2z}
                 svg{M108.9,79.1h41.6c39.6,0,57,28.3,57,53.6c0,27.5-21.5,53.6-56.8,53.6h-41.8V79.1z M124.3,172.4h24.5c34.9,0,42.9-26.5,42.9-39.7c0-21.5-13.7-39.7-43.7-39.7h-23.7V172.4z}
                 svg{M88.7,56.8c0,5.5-4.5,10.1-10.1,10.1c-5.6,0-10.1-4.6-10.1-10.1c0-5.6,4.5-10.1,10.1-10.1C84.2,46.7,88.7,51.3,88.7,56.8z};
  }
}
\newcommand\orcidicon[1]{\href{https://orcid.org/#1}{\mbox{\scalerel*{
\begin{tikzpicture}[yscale=-1,transform shape]
\pic{orcidlogo};
\end{tikzpicture}
}{|}}}}
\usepackage{hyperref} %<--- Load after everything else

\usepackage{tikz}
\usepackage{textcomp}
\usepackage{hyperref}
\usepackage{lipsum}
\ifCLASSOPTIONcompsoc
  \usepackage[nocompress]{cite}
\else
  \usepackage{cite}
\fi
\begin{document}
\title{Cache Where you Want!\\Reconciling Predictability and Coherent Caching}

\author{
  \IEEEauthorblockN{
    Ayoosh Bansal \orcidicon{0000-0002-4848-6850}\IEEEauthorrefmark{1},
    Jayati Singh \orcidicon{0000-0003-1528-7369}\IEEEauthorrefmark{1},
    Yifan Hao\IEEEauthorrefmark{1}, Jen-Yang Wen\IEEEauthorrefmark{1},
    Renato Mancuso\IEEEauthorrefmark{2}
    and Marco Caccamo\IEEEauthorrefmark{3}
  }
  
  \IEEEauthorblockA{
    \IEEEauthorrefmark{1} University of Illinois at Urbana-Champaign,
    \{ayooshb2, jayati, yifanh5, jwen11\}@illinois.edu
  }
  \IEEEauthorblockA{
    \IEEEauthorrefmark{2} Boston University, rmancuso@bu.edu
  }
  \IEEEauthorblockA{
    \IEEEauthorrefmark{3} Technical University of Munich, mcaccamo@tum.de
  }
}

\newcommand\copyrighttext{%
  \footnotesize \textcopyright 2021 IEEE. Personal use of this material is permitted.
  Permission from IEEE must be obtained for all other uses, in any current or future media,
  including reprinting/republishing this material for advertising or promotional purposes,
  creating new collective works, for resale or redistribution to servers or lists,
  or reuse of any copyrighted component of this work in other works.
%   DOI: \href{https://ieeexplore.ieee.org/abstract/document/9134262}{10.1109/MECO49872.2020.9134262}
  }
\newcommand\copyrightnotice{%
\begin{tikzpicture}[remember picture,overlay]
\node[anchor=south,yshift=10pt] at (current page.south) {\fbox{\parbox{\dimexpr\textwidth-\fboxsep-\fboxrule\relax}{\copyrighttext}}};
\end{tikzpicture}%
}%
% *** %

\maketitle

\begin{abstract}
Real-time and cyber-physical systems need to interact with and respond to their physical environment in a predictable time. While multicore platforms provide incredible computational power and throughput, they also introduce new sources of unpredictability. Large fluctuations in latency to access data shared between multiple cores is an important contributor to the overall execution-time variability. In addition to the temporal unpredictability introduced by caching, parallel applications with data shared across multiple cores also pay additional latency overheads due to data coherence.

Analyzing the impact of data coherence on the worst-case execution-time of real-time applications is challenging because only scarce implementation details are revealed by manufacturers.  This paper presents application level control for caching data at different levels of the cache hierarchy. The rationale is that by caching data only in shared cache it is possible to bypass private caches. The access latency to data present in caches becomes independent of its coherence state. We discuss the existing architectural support as well as the required hardware and OS modifications to support the proposed cacheability control. We evaluate the system on an architectural simulator.
We show that the worst case execution time for a single memory write request is reduced by 52\%.
Benchmark evaluations show that proposed technique has a minimal impact on average performance.
\end{abstract}


\IEEEpubidadjcol
\copyrightnotice

\IEEEdisplaynontitleabstractindextext
\IEEEpeerreviewmaketitle

%\IEEEraisesectionheading{\section{Introduction}\label{sec:introduction}}
\section{Introduction}
\label{sec:introduction}

\IEEEPARstart{T}{he}
 last decade has witnessed a profound transformation in the way real-time systems are designed and integrated. At the root of this transformation are the ever growing data-heavy and time-sensitive real time applications. As scaling in processor speed has reached a limit, multi-core solutions~\cite{darksilicon} have proliferated. Embedded multi-core systems have introduced a plethora of new challenges for real-time applications. Not only this adds a new dimension to scheduling, but remarkably the fundamental principle that worst-case execution time (WCET) of applications can be estimated in isolation has been shaken. 

In multi-core systems, major sources of unpredictability arise from inter-core contention over shared memory resources~\cite{sce,cmu_dram}. Memory resource partitioning techniques present a suitable approach to mitigate undesired temporal interference between cores~\cite{colorlock, memguard, palloc}. However, memory resource partitioning is particularly well suited only for systems where data exchange between cores is scarce or nonexistent~\cite{reconciling}. Heavy data processing or data pipelining workloads, on the other hand, are often internally structured as multi-threaded applications, where coordination and fast data exchange between parallel execution flows on different cores is crucial. This exchange is based on shared memory.

Modern platforms generally feature a multi-level cache hierarchy, with the first cache level (L1) comprised of private per-core caches. When multiple threads access the same memory locations, it is crucial to ensure the coherence of different copies of the same memory block in multiple L1 caches. Dedicated hardware circuitry, namely the \emph{coherence controller}, exists to maintain this invariant. Because maintaining coherence requires coordination among distributed L1 caches, it introduces overhead. 

Cache coherence introduces two main obstacles for real-time systems. First, hardware coherence protocols are a preciously guarded intellectual property of hardware manufacturers. As such, scarce details are available to study the worst-case behavior for coherent data exchange. Second, coherence controllers are not designed to optimize for worst-case behavior. One approach to achieve predictable coherence consists of re-designing coherence protocols and controllers~\cite{pmsi, hourglass}. Doing so, however, requires extensive modifications to existing processor design, with a possibly significant impact on performance.

In this paper, we propose a new approach to achieve predictable coherence. The key intuition is that if memory blocks accessed by multiple applications are cached only in shared levels ---e.g., last-level cache--- multiple copies of cache lines do not exist and coherence is trivially satisfied. Based on this, we define a new memory \emph{type} that is non-cacheable in private (inner) cache levels, but cacheable in shared (outer) caches, namely \emph{Inner-Non-Cacheable, Outer-Cacheable (INC-OC)}. INC-OC memory type coexists with traditional memory types. Control over which type of memory should be used for different areas of an application's working set is then provided to the developer and/or compiler. The key \textbf{contributions} of this work can be summarized as follows:

\begin{itemize}
    \item A novel solution for predictable time access to coherent data without variability induced by coherence mechanisms.
    \item Prototype evaluation on an architectural simulator~\cite{gitblind}.
\end{itemize}

\section{Solution Overview}
\label{sec:soln}

The main idea is to allow application developers or compilers to use their knowledge of the application to choose between the trade-offs of worst-case vs. average use access time with fine granularity.
The choice of cacheability determines which level of caches can the data be cached in and which levels of cache data cannot be cached.
Data locations for which strong worst-case latency guarantees are required would be selectively cached in shared levels.
Data coherence is achieved as all accesses go to the same cached copy of such data. Coherence overheads and variability are avoided.

\begin{figure}[htbp]
\centering
\includegraphics[keepaspectratio,width=\linewidth]{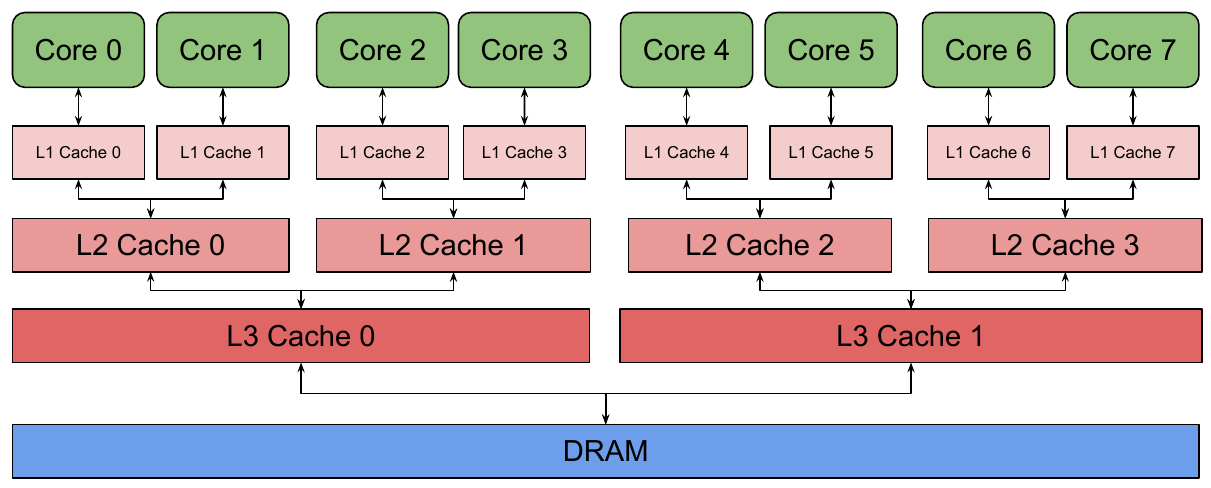}
\caption{\label{fig:multilevel}Generalized multi-socketed multi-cluster processor.}
\end{figure}

Consider a hypothetical system as shown in Figure~\ref{fig:multilevel}.
For each core (C) consider a set M which consists of all memory blocks that are in its data pipeline. A memory block can be a specific cache (\$) or DRAM. For example:

\begin{align}
M(C0) = \{L1 \$ 0,L2 \$ 0,L3 \$ 0,DRAM\} \\
M(C1) = \{L1 \$ 1,L2 \$ 0,L3 \$ 0,DRAM\} \\
M(C7) = \{L1 \$ 7,L2 \$ 3,L3 \$ 1,DRAM\}
\end{align}

We propose that for any cache line which is assessed by only a subset of cores, the cache line be stored/cached on only the intersection of the sets M of these cores. Consider a cache line accessed by Core 0 and Core 1. This cache line can be stored in L2 Cache 0, L3 Cache 0 and DRAM but not in L1 Cache 0 or L1 Cache 1.

\begin{align}
M(C0) \cap M(C1)= \{L2 \$ 0, L3 \$ 0, DRAM\}
\end{align}

Consider another cache line that is accessed from Core 0 and Core 7. Based on our proposal this data block should not be cached at all and be stored in DRAM only.

\begin{align}
M(C0) \cap M(C7)= \{DRAM\} 
\end{align}

Under these restrictions, explicit mechanisms to maintain data coherence are not required. The restrictions should be expressed at application level, possibly as granular as a cache line, so hard real time applications can leverage predictable access time to shared data, while non real time applications running on the same system can enjoy the higher throughput of hardware managed cache coherence as they are not affected by the occasional spikes.

In reality, the closest existing support is cacheability control for two cache levels expressed  at memory page granularity in ARVv8-A ISA, as further described in Section~\ref{sec:memtypes}. The implementation and evaluation are based on an ARVv8-A processor simulator, Figure~\ref{fig:simplesystem}, and limited to two cacheability levels.
\section{Related Work}
\label{sec:relwork}

Multi-core systems have enabled multi-threaded real-time workloads.
Real-time applications with parallel, precedence-constrained processing tasks are often represented as periodic (or sporadic) DAG tasks~\cite{dag-rtss-2012}. Scheduling of DAG tasks has received considerable attention~\cite{dag-rtss-2012, improved-DAG-sched-ecrts14, feasib-analysis-ecrts13}. Alongside, application-level frameworks to structure real-time applications to follow the DAG model have been proposed and evaluated~\cite{rt-openvx-rtss18}.

In practice, however, the implementation and analysis of DAG real-time tasks is challenging because analytically upper-bounding the worst-case execution time (WCET) of a processing node is hard. As multiple DAG nodes are executed in parallel, they interfere with each other on the shared memory hierarchy. Well known sources of interference arise from space contention in the shared cache levels~\cite{origin, survey}; contention for allocation of miss status holding registers (MSHR)~\cite{taming-nb-caches}; bandwidth contention when accessing the DRAM memory controller~\cite{memguard,bw_oslevel}; and bank-level contention due to request re-ordering in the DRAM storage \cite{palloc}.
Our work focuses on tightening and simplifying the analysis of the WCET bound by making shared data accesses immune from unpredictable temporal effects of cache coherence controllers.

Cache contention among parallel tasks is a major cause of interference~\cite{survey}.
Mitigation approaches include selective caching \cite{bypassl2inst,bypassl2data} and cache partitioning \cite{colorlock, Liedtke:1997:OCP:523983.828369}.
Strict resource partitioning among cores is effective for independent tasks. But data sharing is necessary to implement processing pipelines, such as real-time tasks following the sporadic DAG model. In~\cite{mc2}, the authors acknowledge that data-sharing between tasks is inevitable in mixed criticality, multi-core systems.
In light of these considerations, previously proposed solutions have been adapted to allow sharing~\cite{mc2_02, mc2_01}.

The problems introduced by data sharing in real-time signal processing applications were studied in~\cite{giovani, giovani_2, ospert}, which demonstrate that the overhead from cache coherence protocols can severely diminish the gains aimed to be achieved through parallelism in multi-core systems.
But more importantly, the overheads of cache coherence are unpredictable and can have large variance.
In this work we focus exclusively on the unpredictability caused by cache coherence. Our solution is to allow shared data to bypass private levels and be cached directly in shared cache levels.

Previous works have addressed cache coherence in multiple ways.
Giovani et al. propose a coherence aware scheduler \cite{giovani_2} which staggers the execution of tasks that share data. The solution only works at task level granularity and may force idle times on processor.
Predictable MSI \cite{pmsi} solves the coherence unpredictability by using a TDMA coherence bus and a modified MSI coherence protocol. The solution is invisible to software and provides predictability with reasonable overheads, but requires major changes to hardware coherence controllers that are difficult to implement and verify \cite{coherence_verif1, coherence_verif2, coherence_verif3}.
MC2 \cite{mc2} improved upon \cite{mc2_01, mc2_02} by allowing data sharing across processors. The coherence effects are avoided by making the shared memory uncacheable or assigning tasks with shared memory accesses to the same core. The scheduling option is restrictive and like \cite{giovani_2} may force processor idling. Extra accesses to uncached memory, i.e. main memory, are slow and may increase the WCET.
On-Demand Coherent Cache \cite{lockcritical} converts tasks accessing shared data to critical sections, hence disallowing concurrent execution of any tasks that share data.
SWEL \cite{swel} focuses on high performance computing and message passing workloads. It proposes heuristic mechanisms in hardware to cache only private and read only data in L1 caches. There are no predictability assurances as the cache line placements decisions are based on observations at run time and optimize for throughput.

Our proposed solution does not impose any scheduling restrictions on the application and does not burden it to maintain coherence. In addition, it provides the developer  freedom to choose which data to cache where, to optimize the average and worst-case performance trade-offs. This solution can be implemented without any changes to hardware coherence implementations and works with any coherence protocol. Minimal changes to the cache controller's logic are required. The overall effect is a complete avoidance of cache coherence overheads based on the developer's choices.

\section{Background}
\label{sec:background}

Let us first familiarize ourselves with the terms and concepts used in this work.

\subsection{Cache Coherence}
\label{sec:bg_msi}
Cache Coherence~\cite{Patterson:2007:COD:1535294} is a feature of modern multi-level, distributed CPU caches. In traditional cache architectures, it is fundamental that the contents of private levels of caches are kept \emph{coherent} across multiple cores.  A hardware cache coherence controller is present for this purpose. It ensures that any valid copies of a cache line contain the same data.
Cache coherence controllers work by assigning additional states to cache lines. Let us consider the example of \textbf{\textit{MSI}} cache coherence protocol~\cite{1269047} as shown in Figure \ref{fig:msi}. In this protocol a cache line can be in one of the following three states:

\begin{figure}[htbp]
\centering
\includegraphics[keepaspectratio,width=.85\linewidth]{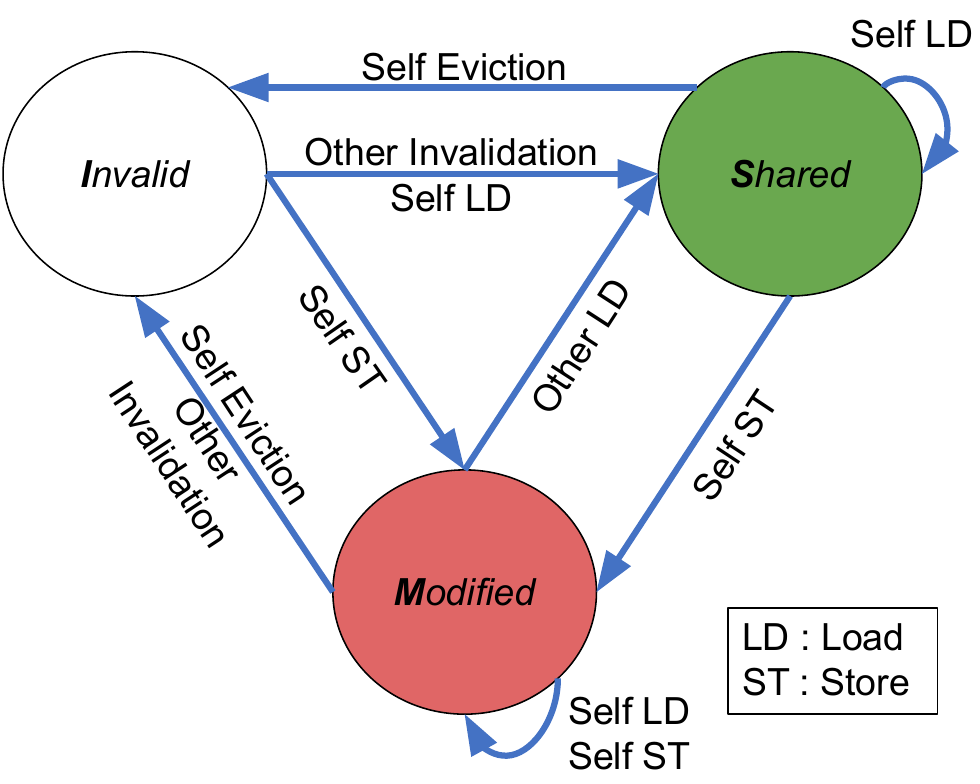}
%\includesvg{msi}
\caption{\label{fig:msi}MSI States and Transitions}
\end{figure}

\begin{itemize}
    \item \textit{\textbf{I}nvalid:} Cache line is not allocated or contains invalid data. This is the initial state.
    
    \item \textit{\textbf{S}hared:} Cache line contains valid data.
    A cache line in shared state can only be read from.
    Other caches may have the same cache line in \textit{\textbf{S}hared} or \textit{\textbf{I}nvalid} state.
    Data cached in this line is the same as the corresponding location in main memory.

    \item \textit{\textbf{M}odified:} Cache line contains valid data.
    A cache line in modified state can be used to read or write data.
    Other caches cannot have the same cache line in any state other than \textit{\textbf{I}nvalid}.
    Cache line contains \emph{dirty} data, i.e., the current content of the cache line may be different from the corresponding location in main memory.
\end{itemize}

A cache line transitions between these state based on \emph{Self} vs. \emph{Other} load/store (LD/ST) events, as shown in Figure~\ref{fig:msi}. Here, \emph{Self} refers events generated by the the core under analysis. Evictions are cache line replacements. \emph{Other} refers to messages to handle events by other cores.

\subsection{Memory Types}
\label{sec:memtypes}

\begin{table*}[hbp]
\centering
\caption{Memory Types}
\label{tab:abbr}
\begin{tabular}{|l|l|l|}
\hline
Name  &  Cacheability  &  Description\\\hline
Normal Cacheable  &  Inner Cacheable, Outer Cacheable  &  Data caching allowed in all caches\\\hline
Uncacheable  &  Inner Non-Cacheable, Outer Non-Cacheable  &  Data caching not allowed\\\hline
\emph{\textbf{INC-OC}}  &  Inner Non-Cacheable, Outer Cacheable  &  Data caching allowed in Shared caches only\\\hline
\end{tabular}
\end{table*}

A vast majority of modern multi-core embedded systems are implemented using ARM architectures.
We focus on the latest major version ARMv8-A,  extensively used in current platforms.
This includes recent versions of Nvidia Tegra, Qualcomm Snapdragon and Samsung Exynos, among others \cite{arm_cores}. There are 100+ mobile and embedded SoC compliant with ARMv8-A Instruction Set Architecture (ISA).
In this architecture, a uniform physical memory address space describes traditional memory resources (e.g. DRAM space), as well as configuration space for on-chip and external devices. In order to adopt the correct caching policy for any given memory region, the hardware allows specifying a set of meta-data, or \emph{memory type}, for each memory page. The memory type specification informs the hardware of how load/store operations within a given memory range should be handled.
Memory type attributes are encoded in each virtual memory page table descriptor.  In setting up virtual memory, the OS is responsible for encoding the correct memory type in the page table entry (PTE) of any portion of memory being accessed.
If virtual memory is disabled, no memory type can be associated with a memory range. This forces the hardware to be conservative and to treat any load/store operation as non-cacheable accesses. 

ARMv8-A standard allows defining two main attributes in the memory type. First, \emph{cacheability}: i.e., whether or not a memory location should be cached or not. There exist two cacheability attributes: \emph{inner cacheability} and \emph{outer cacheability}. If a memory region is marked as inner (resp., outer) cacheable, its content can be cached in the inner (resp., outer) cache levels. What constitutes inner vs. outer is implementation defined. Generally, however, private cache levels (e.g., L1) are inner caches. Conversely, shared cache levels are usually outer caches. 
Second, memory types encode a \emph{shareability} attribute. Once again, this can be specified independently for inner and outer caches. If a memory region is defined as inner shareable, any of its cached lines are kept coherent by the hardware in the inner cache levels. The same goes for outer shareable memory.

In this work, we focus on a subset of possible memory types~\cite{armioc}. Specifically, Table~\ref{tab:abbr} defines the memory attributes used throughout the paper.
The default memory type is \textit{Normal Cacheable}. This type of memory is cacheable (and shareable) at all levels of caches.
The other frequently used memory type is \textit{Uncacheable}. This memory type is typically used to describe I/O memory. We define a new memory type: \textit{Inner Non Cacheable, Outer Cacheable (\textbf{INC-OC})} and also address kernel support in Section \ref{sec:impllinux}. This type of memory is accessed by the processor cores only. It is cached in all shared (outer) cache levels but not cached in any caches private (inner) to any cores.

\subsection{Architectural Support}
\label{sec:arch_support}
One of our goals is minimal changes. So we first explore existing support for cacheability control in popular Instruction Set Architectures (ISA) and corresponding compliant processors.
We find that while ARMv8-A ISA supports inner and outer cacheability control, hardware implementations simplify away this support. Conversely, X86 and MIPS ISA do not support high granularity cacheability control for different cache levels.

\begin{itemize}

%\subsubsection{X86}
\item \textbf{X86} : 
% Section 11.3, 11.11, 11.12 at 
% https://www.intel.com/content/dam/www/public/us/en/documents/manuals/64-ia-32-architectures-software-developer-vol-3a-part-1-manual.pdf
Intel 64 defines various levels of caching like Uncacheable, Write Combining, Write Through, Write Back \cite{manual_intel}.
Two methods are provided to specify the type of caching, namely, Memory Type Range Registers and Page Attribute Table.
The defined caching types do not differentiate between the various levels of caches.

\item \textbf{ARM} : 
\label{sec:bgarm}
% 6.4.1 Behavior for different memory types
% http://infocenter.arm.com/help/index.jsp?topic=/com.arm.doc.100095_0002_03_en/Chunk292957963.html
% 17.6.5.1 Behavior for Different Memory Types
% https://developer.nvidia.com/embedded/downloads
% Tegra X2 (Parker Series SoC) Technical Reference Manual
% Or just refer to the one in Google Drive as download requires an account.
ARMv8-A ISA allows managing the cacheability of Inner and Outer regions independently~\cite{manual_armv8a}.
Most ARM processors however simplify their design by treating Inner Non-cacheable Outer Cacheable type as Non-Cacheable memory. This applies to Cortex-A53 \cite{cache_a53}, Cortex-A57 \cite{cache_a57}, Cortex-A72 \cite{cache_a72}.
Other ARM compliant processors implement similar simplifications.
Nvidia Denver and Carmel architectures ignores the Outer Cacheability attribute~\cite{manual_parker,xavier_l1}.

%\subsubsection{MIPS}
\item \textbf{MIPS} : 
% Table 9.2 at
% https://s3-eu-west-1.amazonaws.com/downloads-mips/documents/MD00091-2B-MIPS64PRA-AFP-05.04.pdf
MIPS treats cacheability control in a simpler manner. MIPS32/64 ISA only defines Cached and Uncached memory types  \cite{manual_mips}.
A lot of fields are left as implementation dependent though.
One of the recent MIPS processors, M6200 supports only Cached and Uncached memory types~\cite{manual_mips_m6200}.

\end{itemize}

\section{Approach Overview and Motivation}
\label{sec:motivation}
The main idea behind our solution is to allow application developers or compilers to use their knowledge of the application to choose between the trade-offs of worst-case vs. average use access time.
The choice of cacheability determines whether or not certain data will be cached in private L1 levels.
Data locations for which strong worst-case latency guarantees are required can be selectively cached only in shared levels.
Data coherence is achieved as only one cached copy of such data locations can exist. Coherence overheads and variability are avoided for any access to such locations.

\subsection{Architectural Prerequisites}
\begin{figure}[htbp]
\includegraphics[keepaspectratio,width=\linewidth]{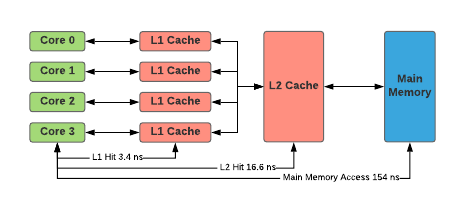}
\caption{\label{fig:simplesystem}System model}
\end{figure}

For cacheability control to provide predictable time access, two core requirements must be met. First, a \emph{shared cache level} must exist between the entities that share the worst case latency sensitive data. Consider Multi-Socket Multi-Core (MSMC) architectures. The chips in such sockets do not have a shared cache level. Their first common memory level is the main memory and hence our solution would translate to using uncacheable memory type. The second requirement is that for the given shared cache level, a \emph{method} must exist to \emph{limit the cacheability} to that level. Consider a multilevel cache architecutre like Nvidia Carmel SoC. Cluster of 2 cores share a L2 cache level. 2 such clusters share an L3 cache. In such cases, either the data sharing needs to be limited or enough memory types must exist to limit cacheability to each cache level. This level of fine grained control is not supported by current ARMv8-A~ISA. For the remainder of this paper we consider a system as shown in Figure~\ref{fig:simplesystem}.

\subsection{Coherence Cost}
\label{sec:coherence_cost}

Cache coherence introduces a new dimension to cache function.
A cache hit can no longer be defined as simply having the data in the cache.
Correct state and privilege are now also a requirement for a hit.
The implementation details of cache coherence controllers are proprietary and hence it is generally difficult to estimate or measure the exact latency of every operation.
Cache access operations can be one of:

\begin{itemize}
    \item \textit{Hit:} The data block is present in the cache and in a state that allows the desired operation. For example \textit{\textbf{S}hared} for Load and \textit{\textbf{M}odified} for Store.
    \item \textit{Miss:} The data block is not present in the cache and needs to be retrieved from a lower cache level or main memory.
    \item \textit{Coherence Miss:} The data block is present in the local cache or a remote cache at the same level. But the state of the block does not allow the desired operation. For example Store on a cache line in \textit{\textbf{S}hared} state.
\end{itemize}

\begin{figure*}[htbp]
\centering
\includegraphics[keepaspectratio,width=.85\linewidth]{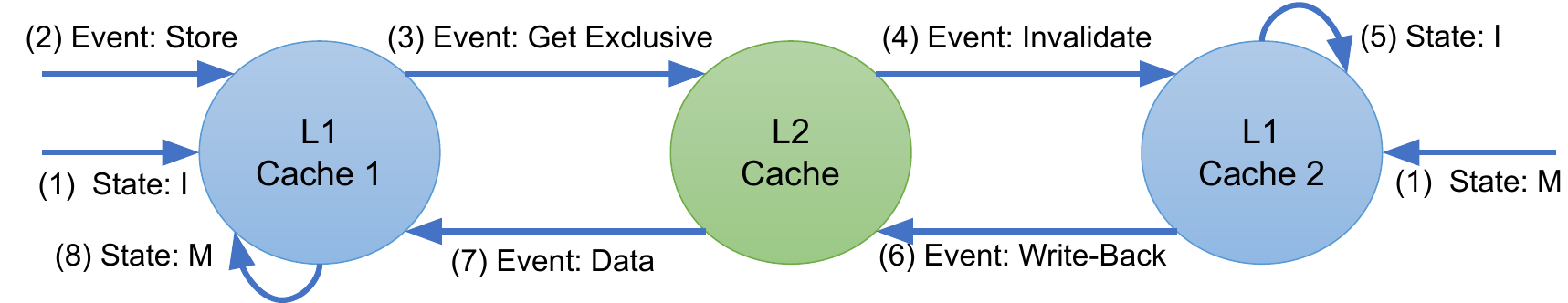}
\caption{\label{fig:dirtymiss}Transitions of a Dirty Miss}
\end{figure*}
    
Another example is shown in Figure \ref{fig:dirtymiss}.
Consider a 2 core processor with 2 cache levels.
In this example, the core attached to L1 Cache1 initiates a store operation. This cache does not have the data block for the Store, but L1 Cache2 has the data block in \textit{\textbf{M}odified} state. Cache2 has to invalidate its cache line and write back the dirty data to the shared L2 cache. The L2 cache can then send the data to L1 Cache1 which can finally execute the Store. L1 Cache1 now contains the cache line in Modified state. These series of events can lead to long latency in executing a single memory access. We refer to this situation as a \emph{Dirty Miss} in this paper.

\begin{figure}[htbp]
\centering
\includegraphics[keepaspectratio,width=.75\linewidth]{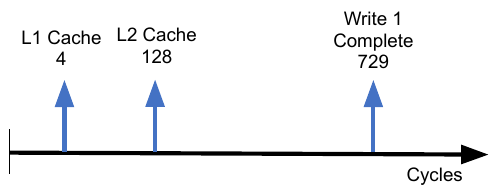}
\caption{\label{fig:dirtymiss_timeline}Timeline of a Dirty Miss}
\end{figure}

Figure \ref{fig:dirtymiss_timeline} illustrates the cost of a \emph{dirty miss}.
Using an instrumented simulation, see Section \ref{sec:trace_mode},  we can set up custom scenarios.
Simulation logs show when some events of interest complete. The time reported is simulation ticks which are a direct equivalent to cycles in a hardware system.

At time 0, a core initiates a single write request. The target cache line is in \textit{\textbf{M}odified} state in another L1 cache.
It takes 4 cycles for the L1 cache to get the request from the processor, determine the current state of the cache line and take actions accordingly.
It takes a further 124 cycles or total 128 cycles for L2 cache to receive the first request and take actions accordingly.
The next 601 cycles are spent in completing the coherence steps, as shown in Figure \ref{fig:dirtymiss}, and delivering data to the requesting L1 cache.
It takes $4.7\times$ cycles to resolve a dirty miss compared to L2 communication delay.

\subsection{Coherence Complexity}
Hardware cache coherence simplifies the development of general purpose multi-threaded software.
Many applications are served well by the transparent handling of data coherence by the hardware. But for real-time applications this creates another uncontrolled source of unpredictability in their worst-case execution time.
Cache coherence protocols in SoCs are defined by vendors with only the main stable states \cite{a53_coherence, xavier_l1}.
There are a plethora of transient states in coherence state machines and many low level details that impact the overall coherence state machine operation \cite{abts2003so}.
This makes any analysis on existing cache coherence controllers difficult.
Our approach of caching shared data in shared cache only, completely removes the cache coherence controller from the shared data access process.
Hence the effect of coherence on worst case analysis and during certification of the system is trivially handled.

\subsection{Private vs Shared Access}
\label{sec:motiv_plat}
In this section we discuss the difference between accessing Shared vs Private data on two real platforms.
\begin{itemize}
    \item Cortex-A53~\cite{a53}: Quad-core ARMv8-A processor~\cite{platform_link}.
    \item Xeon E5-2658~\cite{intelproc}: 14 core, 2 hyperthread per core, Intel processor on a desktop workstation.
\end{itemize}

\begin{figure}[htbp]
\centering \vskip-5mm
\includegraphics[keepaspectratio,width=\linewidth]{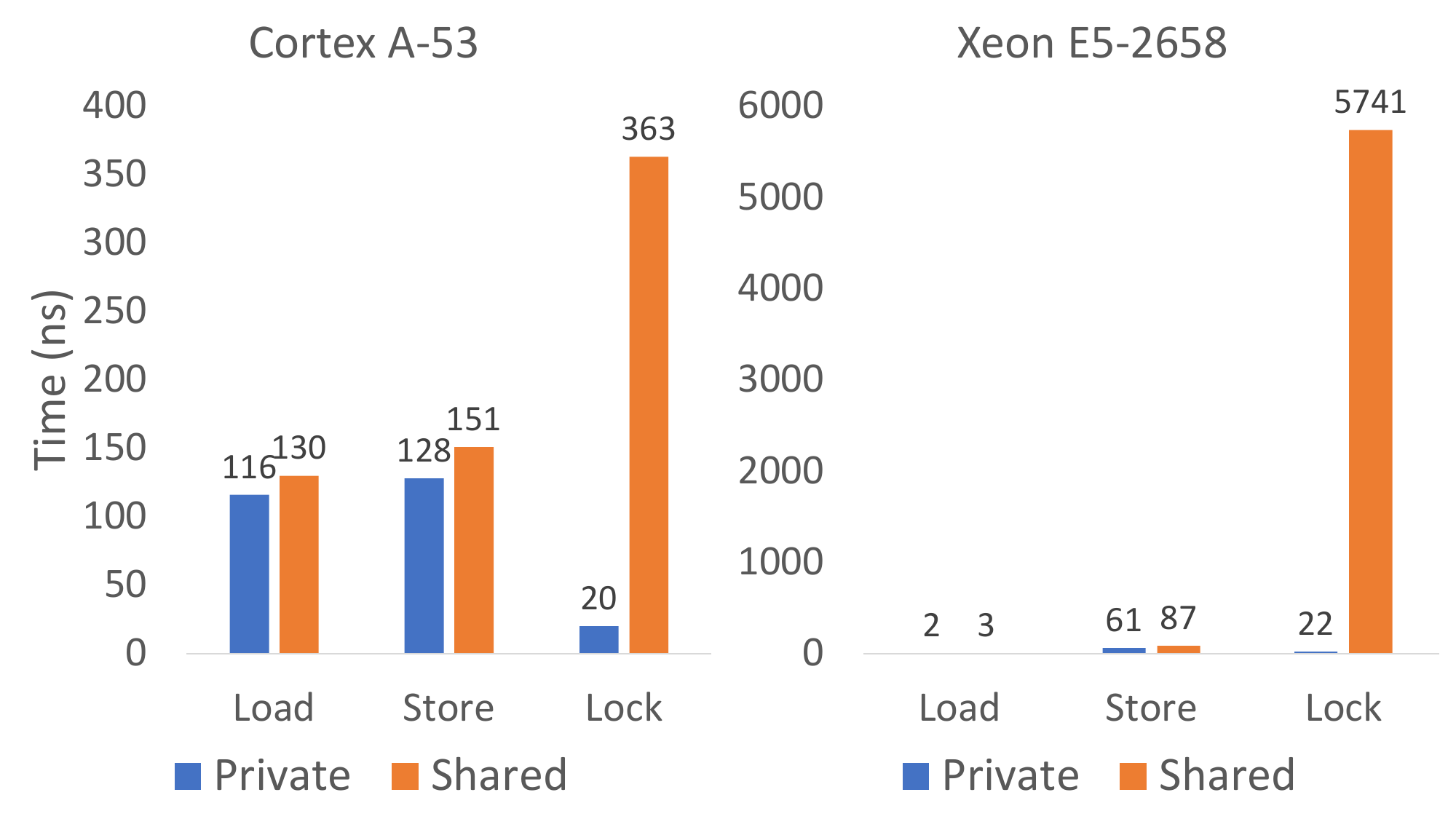}
\caption{\label{fig:latpvss}Private vs Shared Access Latency}
\end{figure}

We developed synthetic benchmarks to study the effect of cache coherence on real platforms. We measure the average latency to complete a Load or Store to data already present in L1 caches. All cores do the same operations simultaneously. The resulting average latency is a combined effect of single access latency, parallelization, bandwidth contention and opportunistic hardware optimization like prefetchers. Consider Figure~\ref{fig:latpvss}. For the first cluster of bars, \emph{Load}, every core reads sequential memory. For the second cluster, \emph{Store}, every core writes to sequential memory and then reads the value back. The read back ensures that the captured latency includes Store to cache and not to an internal write buffer only.
In \emph{Private} case (left bars) every core accesses private data sets, in \emph{Shared} (right bars) they all share one data set. The \emph{Shared} case therefore shows the additional overhead of cache coherence protocols.

%\vskip-5mm

The Lock latency is the average latency to acquire and release a spinlock. In case of \emph{Private} every core accesses a different spinlock.
All lock operations complete within the Private L1 cache and there is no waiting on the lock and no one else is trying to acquire it.
In \emph{Shared} every core acquires and releases the same lock. The critical section is empty. Spinlock implementation is below and uses gcc built-in atomics~\cite{gccprebuilts}. The acquire and release sequence is hence less than 10 assembly instructions. In case of shared lock, hence, most of the time is spent on access contention.

% \begin{minted}
% [
% frame=lines,
% framesep=1mm,
% baselinestretch=.5,
% fontsize=\footnotesize
% ]{c}
% lock: while(__sync_lock_test_and_set (&lck, 1)) {};
% unlock: __sync_lock_release (&lck);
% \end{minted}

\begin{lstlisting}[style=CStyle]
lock: while(__sync_lock_test_and_set (&lck, 1)) {};
unlock: __sync_lock_release (&lck);
\end{lstlisting}

While exact overheads of cache coherence are hard to measure on real platforms, it is evident that the latency to access data is dependent on whether it is being shared across different cores. The Load/Store measurements represent latency differences near full memory bandwidth and hence the effect of coherence itself is diminished. Locks on the other hand require that the underlying micro-ops/instructions complete in order. Locks are hence affected more by the overheads of maintaining coherence.

\section{Implementation}
\label{sec:impl}

As noted in Section \ref{sec:arch_support}, existing COTS ARMv8-A platforms cannot use INC-OC memory type hence the evaluation is limited to simulations. We implement the controlled cacheability on the gem5 architectural simulator~\cite{gem5} to realize a system as shown in Figures \ref{fig:simplesystem} and \ref{fig:impl}.
The simulation system is configured to be representative of a Cortex-A53 \cite{a53}.
The memory hierarchy is comprised of 32 KB L1 cache per core, 2 MB L2 cache and 4 GB DRAM. This is in line with typical Cortex-A53 based SoCs~\cite{platform_link}.
We chose configurable cache parameters to mirror Cortex-A53 though some differences would surely exist. For both the Cortex-A53 \cite{a53,platform_link} and the simulated system, memory access latency are as shown in Figure~\ref{fig:simplesystem}.

Further in this section we describe our implementation
and how it would function on a real system. Our modifications add the support
for the  INC-OC memory type in gem5 simulator.
The design is presented top to bottom, starting from application layer and all the way to cache microarchitecture. 
This design is close to what the authors, to the best of their knowledge, believe a hardware implementations should be like.
gem5's cache framework \emph{Ruby} does not support uncacheable memory type. Access latency to uncacheable memory is much higher than any cache. Hence any comparison with uncacheable memory is not interesting.

\subsection{Application}
\label{sec:implapp}
Our primary aim is to provide applications the mechanism to decide between the tradeoffs of worst case vs average access time.
ARM ISA allows expressing cacheability of memory at a per-page granularity.
We modified \textit{mmap}~\cite{mmap} memory allocation API to accept additional flags that are used by the kernel to determine the  cacheability of allocated pages.
As part of the evaluation of this work we modified some standard benchmarks to use this cacheability control. 
An example INC-OC allocation is shown here:

% \begin{minted}
% [
% frame=lines,
% framesep=1mm,
% baselinestretch=.5,
% fontsize=\footnotesize
% ]{c}
% buf = mmap(0, size, PROT_READ | PROT_WRITE,
%             MAP_SHARED | MAP_INCOC, fd, offset);
% \end{minted}

\begin{lstlisting}[style=CStyle]
buf = mmap(0, size, PROT_READ | PROT_WRITE,
            MAP_SHARED | MAP_INCOC, fd, offset);
\end{lstlisting}

\subsection{Kernel}
\label{sec:impllinux}

For an OS to provide the cacheability control to the userspace application, two components are required: first, APIs to allow applications to choose the memory type, as shown above. Second, page table entries need to be set up with the right value as defined by the ISA to use the INC-OC memory type.
We implemented both these components for Linux ARM64.
Our \textit{mmap} syscall implementation allows for additional flags to be passed by an application.
These flags are then used to determine the memory type to set the page tables.
We created a new memory type in the kernel for INC-OC.
Linux kernel defines 6 memory types\footnote{arch/arm64/include/asm/memory.h}.
Two more memory types can be defined. So we were able to add the new memory type with minimal changes.
The kernel changes can run on any ARMv8-A compliant platform and set the memory type bits for INC-OC as defined in the ARMv8-A ISA, but the eventual handling depends on the underlying hardware.
In accordance to Cortex-A53 documentation \cite{cache_a53} when such cacheability bits for INC-OC type are set by the Linux kernel, Cortex-A53 treats these memory pages as uncacheable.
This is a simplification in processor implementation as described in Section \ref{sec:bgarm}.
In our simulation system the same cacheability bits allow the underlying caches to handle cacheability of memory requests in accordance to the full specification of the ARMv8-A ISA.
The kernel changes in form of a patch are available~\cite{gitblind}.

\subsection{Processor}
\label{sec:implproc}
We use the existing Timing Simple Processor in gem5 that runs ARMv8-A ISA.
We modify the Translation lookaside buffer (TLB) to cache the additional page table attributes and add them to every memory access sent to the memory subsystem.
Similar changes will be required in the real processors to propagate the cacheability information from page tables to memory requests.

\subsection{L1 Cache Controller}
\label{sec:impll1}

\begin{figure*}[htbp]
\centering
\includegraphics[keepaspectratio,width=.7\linewidth]{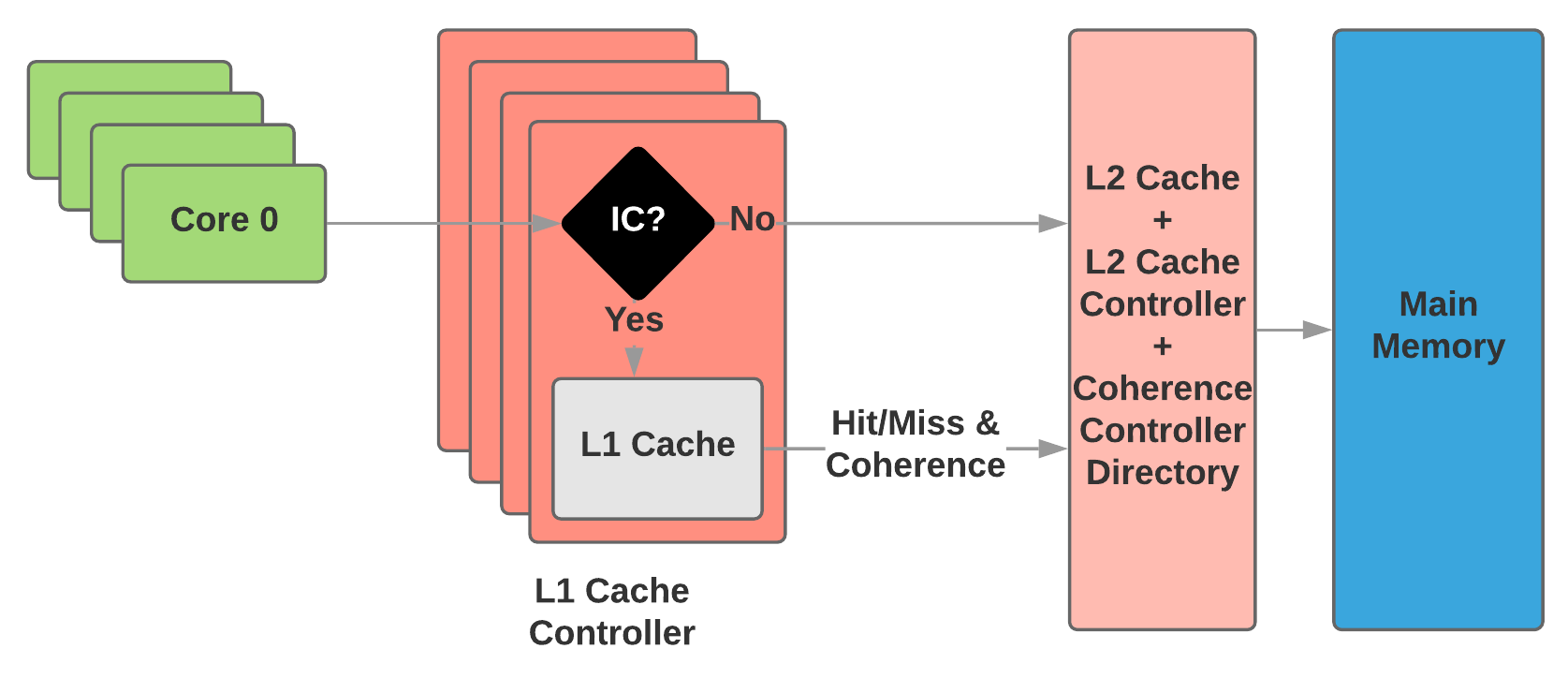}
\caption{\label{fig:impl}Memory Access Flow}
\end{figure*}

We modified the L1 cache controller to check if the request is for Inner Cacheable (IC) memory. The further handling is dependent on this check.

\subsubsection{Normal Memory}
Our L1 cache controller follows the MSI protocol.
The L1 cache controllers do not communicate with each other.
Any L1 cache only communicates with its processor core and the shared L2 Cache.
Hence L1 caches send requests to the L2 cache controller which completes all relevant actions before sending a response back to the L1 cache.

\subsubsection{INC-OC Memory}
If a memory request is marked as Inner Non-Cacheable the request is directly forwarded to the L2 cache as shown in Figure \ref{fig:impl}.
Similarly, any responses to these requests from the L2 cache are forwarded to the processor.
Since INC-OC data blocks are never cached in the private L1 caches, multiple copies of the same data can not exist, therefore, no extra logic is required to maintain coherence.
The cost of this change is that all requests to INC-OC cache lines go directly to the L2 cache, including what could have been L1 cache hits.

\subsection{L2 Cache Controller}
\label{sec:impll2}
The L2 cache is the only shared cache in our system as depicted in Figure~\ref{fig:simplesystem}.
It is strictly inclusive i.e. it contains any cache line that is cached in an L1 cache. The DRAM subsystem that connects to this L2 cache was not modified and emulates a constant access time main memory.

\subsubsection{Normal Memory}
The L2 cache controller serves as the coherence directory for this machine, in addition to the shared cache.
It maintains information about all cached lines and manages all requests from L1 caches.
It is able to ascertain the steps involved in fulfilling such requests, like sending out invalidations or collecting acknowledgements for the invalidations.

\subsubsection{INC-OC Memory}
L2 Cache Controller was modified to support the INC-OC memory type.
INC-OC cache lines share the same memory space as normal memory type.
But once allocated as an INC-OC cache line, the coherence state machine marks them separately and treats them like single-core system cache lines.
A cache line can convert between INC-OC or normal types but needs to be invalidated in between.
A cache line cannot be simultaneously treated as INC-OC and Normal Cacheable memory.
Due to the direct forwarding of all INC-OC requests to L2 caches, there is increased contention on the L2 cache bandwidth.
Since the L2 cache is already inclusive of L1 caches there is no increase in contention for L2 cache space.
This change converts the coherence problem to a cache bandwidth contention problem which is well studied in literature~\cite{memguard,bw_oslevel}.

A major functional change in the L2 cache is the requirement to handle Load/Store exclusive (LDXR/STXR) instruction pair. To this end we maintain a markup of LDXR instructions on cache lines and only accept the STXR if the cache line has not been modified since the corresponding LDXR from the same core. This is a new requirement for INC-OC memory type, as in case of Normal memory type, all LDXR/STXR instructions are handled in the private cache itself.

\subsection{Simulation Modes}
We use the simulation system in two modes. Trace mode uses the memory subsystem in isolation to replay data access traces. Full system mode emulates a complete hardware platform.

\subsubsection{Trace Mode Simulation}
\label{sec:trace_mode}
\emph{Trace} mode for gem5 was developed by Hassan et al. \cite{pmsi}.
It allows using the gem5 memory subsystem alone.
The processor subsystem is replaced by a synthetic request injection.
Based on an input trace file a dummy processor generates memory requests.
The trace file contains memory address, operation (read/write) and time stamp for initial request injection.
We manually construct traces to create custom scenarios. These scenarios do occur in real systems but are difficult to reproduce and observe.

\subsubsection{Full System Simulation}
\label{sec:implfs}
In this mode the simulator emulates a real platform. The cache and memory subsystems are the same as previous mode. But there is a full fledged ARMv8-A compatible processor subsystem.
This mode supports running a Linux kernel. Benchmark applications are oblivious of the underlying simulator and run as if on a real platform.
As noted in Section~\ref{sec:impl} and Figure~\ref{fig:simplesystem}, we have selected simulation parameters so that the simulated system's cache access latency are close to Cortex-A53 \cite{a53,platform_link}.
This full system simulation brings together all aspects of our implementation as described in Sections~\ref{sec:implapp}-\ref{sec:impll2}.
The simulator with all the changes is available~\cite{gitblind}.

\subsection{Discussion and Limitations}
The design is close to what the authors, to the best of their knowledge, believe a hardware implementations should be like.
It seems impractical to have dedicated ports from the processor to each level of the memory system.
So the processor should only interact with the respective L1 caches.
The L1 cache, based on parameters in the memory request, can then determine if it should allocate lines for such a request.
This design choice does slow down access to INC-OC lines. Additional cycles are spent in competing with any pending L1 cacheable memory requests and going though the L1 cache controller logic. 
A direct port would have bypassed all this but we believe that the silicon and metal costs will be prohibitive.
The cascading communication path also makes our solution scalable to multilevel caches as long as there is enough ISA support to express the memory types that define cacheability at each level.

\section{Evaluation}
\label{sec:eval}

The evaluation is based on custom scenarios, microbenchmarks and benchmarks from the SPLASH2 \cite{splash2} suite. 

\subsection{Trace Mode Simulation}
As described before in Section~\ref{sec:trace_mode}, \emph{Trace} mode synthetically injects accesses to the memory subsystem of gem5, based on input trace files.
The traces can be manually written to create specific scenarios.

\begin{figure*}[htbp]
\centering
\includegraphics[keepaspectratio,width=.8\linewidth]{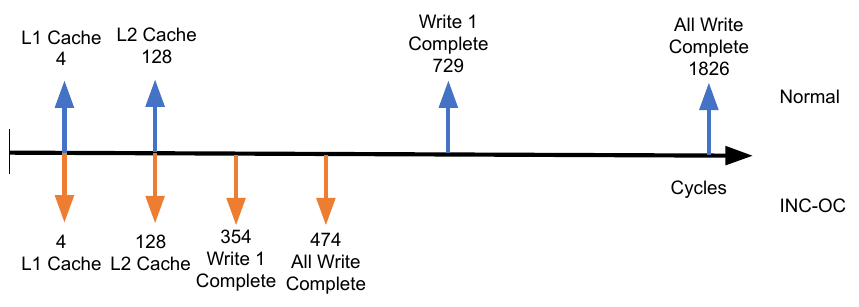}
\caption{\label{fig:wc_both_hit}Worst-Case Write-Request Contention}
\end{figure*}

Figure \ref{fig:wc_both_hit} shows an extended scenario similar to Section \ref{sec:coherence_cost}. A write request to the same address is generated by each core, simultaneously. Write 1 complete is the same \emph{dirty miss} scenario as Figure \ref{fig:dirtymiss_timeline}. We additionally note the time when all 4 write requests finish.
Similar experiments for read requests were conducted but are not shown. \textbf{\textit{S}hared} state data, that can safely co-exist on multiple private caches, is served better by normal memory than INC-OC.

The trace files and gem5 coherence log that provides the timing information is available~\cite{gitblind}.

\textbf{\emph{Observation:}} The total time to process the \emph{dirty miss} is \textbf{52\%} shorter for INC-OC memory. For all 4 write requests the total time was reduced by \textbf{74\%}.

\textbf{\emph{Limitation:}} A \emph{dirty miss} is a fairly common occurrence for shared data accesses but multiple parallel write requests for the same data does not happen in well written programs. From a coherence perspective, a similar situation can occur when atomic accesses, like spinlocks, request exclusive access to cache lines in course of executing LDXR instruction.

\subsection{Full System Simulation}
\label{sec:eval_fs}
In full system simulation, programs are run inside a full fledged Linux kernel running over the simulation platform. This presents a close equivalence to a real machine. To reduce the variability in program execution time caused by CPU and DRAM, we use the timing simple models for both included in gem5. These models provide constant time DRAM accesses and simplified instruction pipelines, while still maintaining detailed timing of events. Also, between \emph{Normal} and \emph{INC-OC} only an \emph{mmap} argument flag\footnote{As shown in Section \ref{sec:implapp}.} is changed before compilation. The programs are otherwise identical. All this helps limit the evaluation to purely a comparison between \emph{Normal} and \emph{INC-OC} memory types.

\subsubsection{Synthetic Benchmarks}
\label{sec:eval_fs_micro}

\begin{figure}[htbp]
\includegraphics[keepaspectratio,width=\linewidth]{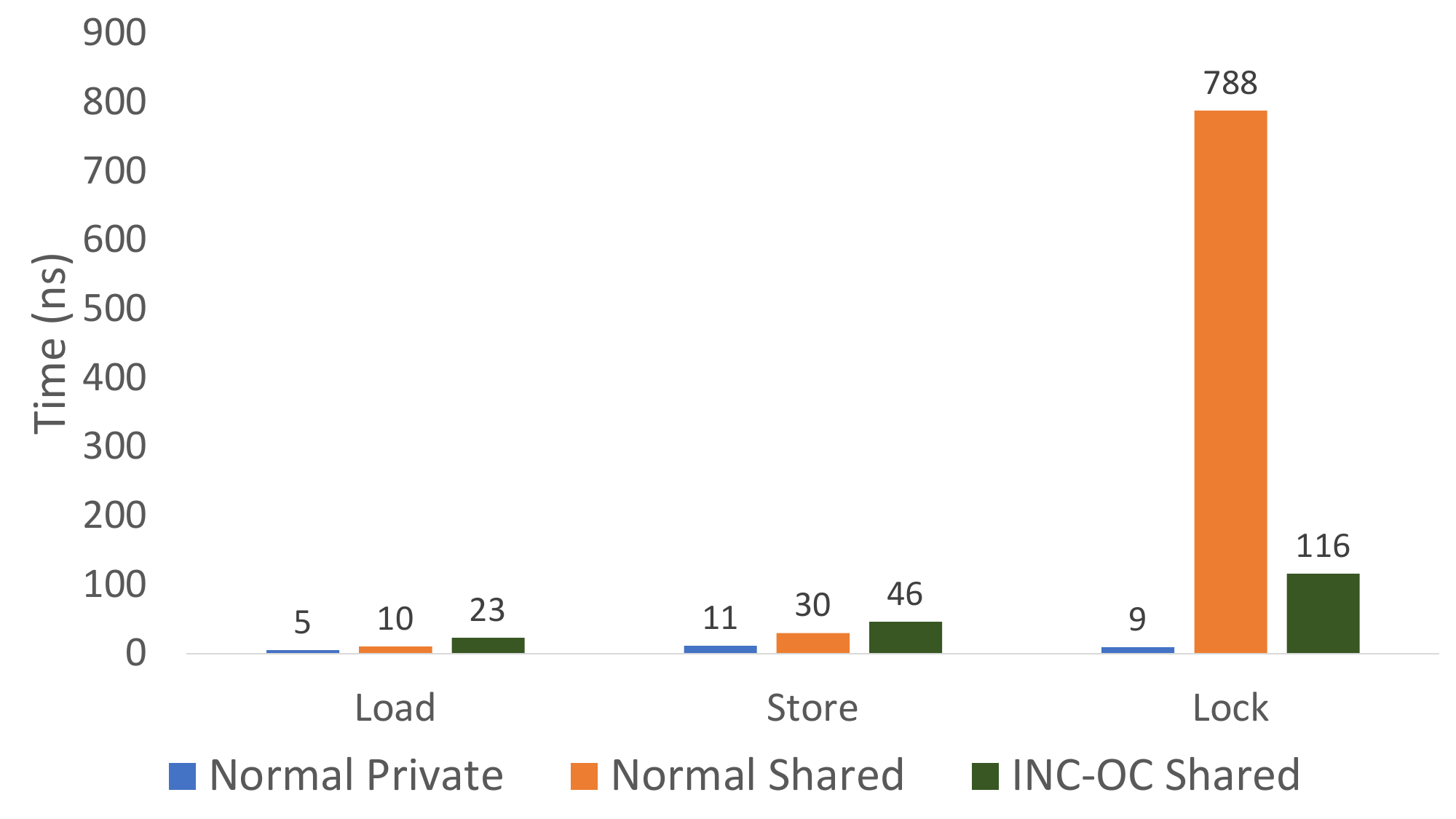}
\caption{Full system simulation synthetic benchmarks}
\label{fig:fs_micro_sim}
\end{figure}

In Figure \ref{fig:fs_micro_sim} we show the results of running the microbenchmarks discussed in Section \ref{sec:motivation}. As before,
Load, Store and Lock refers to the average latency of completing these operations. For Normal Cacheable memory we measure the latency on Private memory blocks per core and also memory blocks shared among all cores. Next, we repeat the shared memory experiments with INC-OC memory type.

\textbf{\emph{Observation:}} 
Latency to acquire locks reduces significantly by the use of INC-OC memory type. Load and Store time is increased for INC-OC.

\textbf{\emph{Discussion:}}
The forced ordering of locking primitives avoids other effects, but since data coherence among contending cores depends on coherence hardware, the effect of coherence dominates. In case of loads and stores the combined effect of bandwidth limitations, additional latency and parallel handling of coherence of individual lines makes INC-OC average access latency higher.

\subsubsection{Benchmark Evaluation}
\label{sec:eval_fs_macro}
\label{sec:opt}

\begin{figure}[htbp]
\includegraphics[keepaspectratio,width=\linewidth]{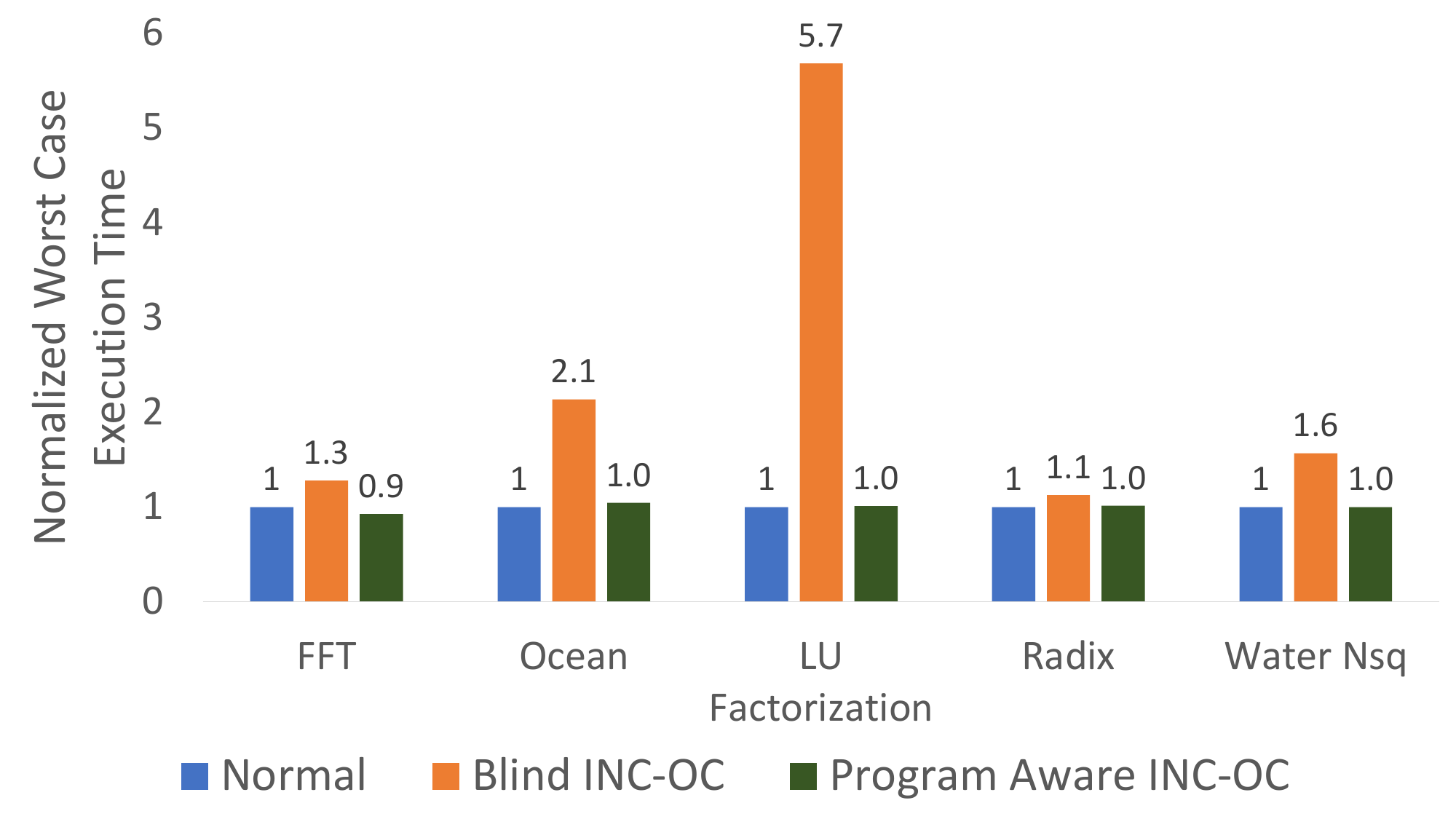}
\caption{\label{fig:opt}Benchmark evaluation for Full System simulation. }
\end{figure}

In Figure \ref{fig:opt} we compare the worst case run time for SPLASH2 \cite{splash2} benchmarks on the full system simulation.
For \emph{Normal} all memory used is normal cacheable.
\emph{Blind INC-OC} blindly allocates every variable, that is visible across threads created by the benchmark, with INC-OC memory type.
For \emph{Program Aware INC-OC} we identify program variables and memory locations that can be safely accessed as Normal memory.
Our optimization treats as normal memory those variables  that (1) are accessed by a single thread, that (2) remain constant in parallel parts of the benchmark, or that (3) are within memory ranges that are divided among threads by offset ranges.
Results are normalized to the \emph{Normal} case. The reported measurements represent the worst-case runtime observed in 100 runs with warmed caches. All benchmarks use spinlocks for synchronization which are allocated with an INC-OC memory type, except in the \emph{Normal} case.

\textbf{\emph{Observation:}}
\emph{Blind INC-OC} execution times are up to \emph{5.7$\times$} higher than \emph{Normal}, while \emph{Program Aware INC-OC}'s performance is near identical as \emph{Normal}.

\textbf{\emph{Discussion:}} The blind approach is overly conservative in INC-OC allocation, while \emph{Program Aware INC-OC} precisely targets truly shared memory requests.
But this is a manual process of understanding the program and making choices about cacheability. We expect this to become a part of the regular practice for the development of multi-threaded real-time applications. Nevertheless, this is the true cost of the proposed solution.
The near identical performance is attributable to a small percentage of INC-OC accesses in \emph{Program Aware INC-OC}, max 0.08\% across benchmarks. This highlights the need for a precise and selective mechanism for handling coherence overheads in multi-threaded and parallel applications\footnote{See Section \ref{sec:selective} for further discussion on this.}.

Due to limitations of space a variability analysis is not shown. gem5 simulator aims at deterministic repeatability of execution. Consequently, even in full system mode the observed standard deviation in execution time of benchmarks is quite small, roughly 1\% of execution time.

\section{Security Implications}
\label{sec:security}

Since we are expanding the user space software's control of low level hardware, security implications need to be carefully considered. 

\subsection{Locks}
RISC and CISC architectures handle atomic operations differently at architectural level. This difference has important implications for the use of INC-OC memory type.

\subsubsection{CISC}
\label{sec:seclockscisc}
Atomic operations in X86, like XCHG, disallow any access to the targeted memory area till all involved micro-operations are complete.
This is achieved by asserting a LOCK~\cite{manual_intel} signal to block any access to the relevant memory buses.
If the targeted memory area is in a private cache the memory location is modified internally and cache coherency mechanism ensures that the operation is carried out atomically.
For atomic operations to INC-OC and Uncacheable memory the LOCK signal will have to be asserted to shared memory buses.
This opens up the possibility of a core starving all other cores of memory bandwidth by continuously executing atomic operations on INC-OC or Uncacheable memory~\cite{x86ulocksecurity}.

\subsubsection{RISC}
RISC ISA like ARM and MIPS, use Load-Linked Store-Conditional semantics to implement lock free atomic operations.
Since system resources are never locked, the vulnerability, as discussed above, does not apply.

\subsection{Cache Conflicts}
As a result of the introduction of INC-OC memory type userspace applications can now directly allocate cache lines in shared cache levels. 
Such allocations can be used to force cache lines owned by other applications to evict from the cache.
But all related attack are already possible with \emph{Normal} cacheable memory via directed private cache misses~\cite{bernstein2005cache}.
The access patterns and size required to achieve the same effect via normal cacheable memory depends on the inclusivity and allocation policy of the caches.

\section{Discussion}
\label{sec:limitations}
In this section we discuss the strengths and limitations of cacheability control and INC-OC memory type.

\subsection{Hardware/Software changes}
Our solution requires changes to existing hardware.
But the changes are minimal as we leverage existing ISA features.
In comparison, competing techniques modify the cache coherence controller implementation and behavior.
Cache coherence protocols are difficult to verify~\cite{1011412} and certify for real-time behaviour.
Any solution that changes the coherence protocol states, transient states or timing etc is hence difficult to adopt.
The development and verification costs can be prohibitive.
INC-OC memory does not modify the coherence protocols or controllers.
The memory type itself classifies the memory requests for INC-OC type to be handled outside of the standard coherence state machine.
So while cache controllers need to be modified to handle the new memory type, the coherence controller itself remains unchanged.

The kernel modifications required to support the INC-OC memory types are also small.
Since Linux kernel already has a notion of \emph{memory types} the introduction of INC-OC type required less than 10 new code lines.
Allowing \emph{mmap} system call to set the memory type in a simple implementation required less than 100 new code lines.

\subsection{WCET Analysis}
Due to the complex nature of the coherence protocols, a WCET analysis of coherence controllers, even for those designed for real-time systems, can be an onerous task. INC-OC memory type eliminates the need for such an analysis.

\subsection{Precise Impact}
\label{sec:selective}
A problem with existing solutions for predictable cache coherence is that they impact all data accesses, independently from the data being private or shared.
On the other hand, INC-OC memory type is used explicitly and precisely.
Default memory type for memory allocation API is normal memory.
As discussed before, the mechanism of this selection can be as simple as passing an additional flag during memory allocation.
The solution can be selectively applied by the developer who can judiciously decide between the worst case and average memory access times on a case by case basis.
For this reason, significant performance benefits can be observed via OC-INC on the same benchmarks compared to the alternative solution proposed in~\cite{pmsi}, namely Predictable MSI (PMSI). Specifically, although~\cite{pmsi} uses a less realistic
simulation setup, applications using PMSI incur a $1.45\times$ slowdown compared to the baseline. Conversely, our evaluations indicate that INC-OC has near identical performance to the \emph{Normal} baseline.
As previously mentioned, this precision comes at the cost of manual application code refinement. We argue however that such refinement could largely be automated at compile-time. That is, as long as identification of shared data can be automatically performed, which has been successfully attempted in the past~\cite{bypassl2data}.

\subsection{Applicability}
Detailed cache coherence parameters of a system may limit if INC-OC memory type is useful.

\subsubsection{Coherence Protocol}
A prerequisite for INC-OC memory type being useful is that the L2 cache access time is less than the worst case access time for normal cacheable memory.
While in our observation that is true for some platforms, it is possible to build a processor where this is not true.
A larger number of cores trying to modify the same data at the same time will increase the worst case if the data is cacheable in private caches.
On the other hand, coherence protocols like MOESI that allow dirty data to be communicated directly among private caches without requiring write-back would decrease the worst case latency for private caches.
The exact applicability of INC-OC memory type depends on the full characteristics of the cache and coherence controllers that are usually not publicly documented by vendors.
Vendors tend to include only the stable coherence state information in their product documentation~\cite{manual_intel}.

\subsubsection{Bus Structure}
We implement a directory based coherence, i.e. where the directory, L2 cache in this case, maintains the list of all L1 caches that are using a particular cache line.
The directory sends directed message to all L1 caches when required.
A snooping coherence bus may reduce the worst case access time for normal cacheable memory, albeit it is known that snooping approaches do not scale well with large number of cores.

\subsubsection{Non-Uniform Cache Architectures}
Non-Uniform Cache Architectures (NUCA)~\cite{nuca} use a physically distributed last level cache to reduce wire delays for cache access.
From a given processor core, different banks of the NUCA cache have different access latency.
For worst case analysis with INC-OC memory type the largest latency bank, accessible by a core, should be considered.

\subsection{Other Sources of Variability}
Use of INC-OC memory type eliminates the inter-core interference due to cache coherence only.
Both shared and private caches are still affected by cache misses and bandwidth contention. 
caused by different cores.
One core's cache line can still be evicted by other core's accesses.
Additionally, the multiple cores still share the memory bandwidth of shared caches and all the way to the memory.
This is another source of contention that the use of INC-OC memory does not remove.
In fact due to direct forwarding of requests to shared caches, the use of INC-OC memory increases the shared cache bandwidth usage.
These sources of contention have existed since the use of memory caches.
They are not specific to multi-core systems.
Many existing works have addressed these problems as discussed in Section \ref{sec:relwork}.

\subsection{Application Models}
\label{sec:appmodel}
The impact of INC-OC memory type is heavily dependent on application characteristics. On the one end of the spectrum are \emph{Non-Blocking algorithms}~\cite{greenwald1999non} or \emph{Worker Queue} models that severely limit data sharing. For these applications INC-OC's contribution will be small. On the other end, \emph{Data Streaming} models or \emph{chains of producers and consumers} involve multiple threads continuously sharing and modifying large amounts of data. In this case, OC-INC would lead to significant improvements in predictability. Due to limitations of platform and benchmarks we have not been able to evaluate these application models yet.

\section{Conclusion}
\label{sec:conclusion}

In this paper we present the INC-OC memory type and a series of mechanisms to select memory types from user-space.
Memory types can be defined at a page granularity. A developer can selectively and judiciously decide which memory type to use based on application requirements.
INC-OC memory type bypasses private caches, hence avoiding coherence overheads and reducing worst-case memory access latency. Overall, judicious use of INC-OC memory type can help applications reduce their worst-case execution-time by reducing the unpredictability arising from black-box hardware coherence management.

\ifCLASSOPTIONcompsoc
  \section*{Acknowledgments}
\else
  \section*{Acknowledgment}
\fi

The material presented in this paper is based upon work supported by
the Office of Naval Research (ONR) under grant number N00014-17-1-2783 and by
the National Science Foundation (NSF) under grant numbers CNS 1646383, CNS 1932529 and CNS 1815891.
M. Caccamo was also supported by an Alexander von Humboldt Professorship
endowed by the German Federal Ministry of Education and Research. Any
opinions, findings, and conclusions or recommendations expressed in
this publication are those of the authors and do not necessarily
reflect the views of the sponsors.

\ifCLASSOPTIONcaptionsoff
  \newpage
\fi

% \IEEEtriggeratref{8}
% \IEEEtriggercmd{\enlargethispage{-5in}}

\bibliographystyle{IEEEtran}
\bibliography{ref}

%\input{bios.tex}

\end{document}